\documentclass[twocolumn,american,english,aps]{revtex4-1}
\usepackage[T1]{fontenc}
\usepackage[latin9]{inputenc}
\usepackage{color}
\usepackage{textcomp}
\usepackage{amstext}
\usepackage{graphicx}
\usepackage{babel}
\begin{document}

\title{Theoretical investigation of an in situ k-restore process for damaged
ultra-low-k materials based on plasma enhanced fragmentation }

\author{Anja F{\"o}rster$^{*,\dagger}$, Christian Wagner$^{\#}$, Sibylle Gemming$^{+,\$,\dagger}$,
J{\"o}rg Schuster$^{*}$}

\affiliation{{*} Fraunhofer ENAS, Technologie-Campus 3, 09126 Chemnitz, Germany}

\affiliation{\# Center for Microtechnologies, TU Chemnitz, Reichenhainer Str.
70, 09126 Chemnitz, Germany}

\affiliation{+ Department of Physics, TU Chemnitz, Reichenhainer Str. 70, 09126
Chemnitz, Germany}

\affiliation{\$ Helmholtz-Zentrum Dresden-Rossendorf, Institute of Ion Beam Physics
and Materials Research, Bautzner Landstra{\ss}e 400, 01328 Dresden, Germany}

\affiliation{$\dagger$Center for Advancing Electronics Dresden, TU Dresden, 01062
Dresden, Germany}

\begin{abstract}
We present theoretical investigations of a k-restore process for damaged pourous ultra-low-k (ULK) materials. The process is based on plasma enhanced fragmented silylation precursors to replace k-value damaging, polar Si-OH and Si-H bonds by k-value lowering Si-CH$_{3}$ bonds. We employ density functional theory (DFT) to determine the favored fragments of silylation precursors and show the successful repair of damaged bonds on our model system. This model system consists of a small set of ULK-fragments which represent various damaged states of ULK materials. Our approach provides a fast scanning method for a wide variety of possible repair reactions. Further, we show that oxygen containing fragments are required to repair Si-H bonds and fragments with dangling Si-bonds are most effective to repair polar Si-OH bonds.
\end{abstract}
\maketitle

\section{Introduction}

The ongoing miniaturization process in the microelectronic industry \citep{01} has led to the use of ultra-low-k (ULK) dielectrics in the on-chip interconnect system. They owe their low dielectric constant (k-value) to pores \citep{02,03,04} and terminal hydrophobic organic species \citep{05} such as methyl groups. During the manufacturing process of integrated circuits (ICs), the k-value of the ULK materials increases. This is due to the loss of methyl groups which are replaced by hydrophilic, polar groups generated by active radicals and highly energetic vacuum-ultra-violet photons that break $\textrm{Si-CH}{}_{3}$ bonds \citep{06}. In particular, Si-H bonds are formed (H-damage), which can create Si-OH bonds (OH-damage) in air contact \citep{07}. Thus, it is necessary to replace the damaging, polar bonds by unpolar ones to decrease the k-value again.

UV assisted thermal curing, the silylation process and the k-restore via hydrocarbon plasma were all tested as possible repair mechanisms \citep{06}. Neither of the three repair mechanisms is ideally suited for porous ULK materials. For example, the UV assisted
thermal curing can lead to the compression, if not the collapse, of porous materials \citep{08}. This is due to the loss of SiOH groups and water molecules without replacements that re-stabilize the porous structure. Further, temperatures of about 600-1000\textdegree C are necessary to completely cure the damage \citep{06}, and thus incompatible with ULSI-CMOS technology. 

The silylation process \citep{09,10,11,12,13}, on the other hand, works at much lower temperatures (T<300\textdegree C). However, due to the size of the silylation precursors they are not able to diffuse into deeper regions of the damaged material. Thus, the restore of the k-value is limited to the surface \citep{14}. 

The third repair process uses hydrocarbon plasma. Small methane fragments are able to diffuse into the pores. However, experimental investigations show that the methane plasma fragments build a carbon rich layer on the surface \citep{15}. While this layer protects the ULK material from further damage \citep{16}, it does not repair the already existing damage. Therefore, an alternative in situ k-restore process is needed: The plasma repair process.

The main concept of this repair mechanism is to perform a plasma enhanced fragmentation of large silylation precursors (6-9  {\AA}). The small fragments (2.5-5  {\AA}) can diffuse deeper into the pores than the silylation precursors do, while maintaining their good repair behavior. The repair temperature has to be low (T<100\textdegree C) to protect the material from thermal damage.

\begin{figure}[b]
\includegraphics{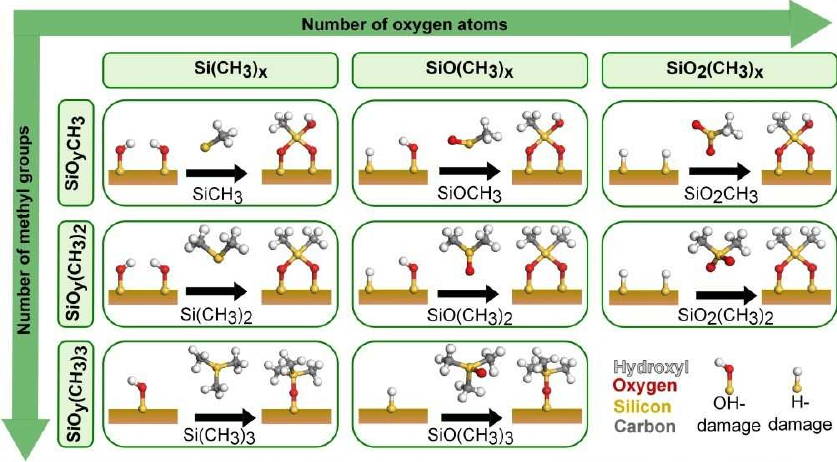}

\protect\caption{(Color online) Overview of possible repair reactions and all considered repair fragments.\label{fig:Overview-repair-reactions}}
\end{figure}

We show an overview of the considered repair fragments and their assumed repair reactions in figure \ref{fig:Overview-repair-reactions}. It is visible that the repair fragments must contain at least one oxygen atom to cure an H-damage. To cure an OH-damage, the silicon atom of the repair fragments must possess at least one dangling bond. Depending
on the number of oxygen atoms and dangling silicon bonds, the repair fragments are able to cure either an OH- or H-damage, or both damage types. Further, the repair molecules that contain exactly one methyl group are prone to further damage after the repair process, while the repair fragments with three methyl groups can only repair one damaged site.

\medskip{}

\section{Computational details}

We employed density functional theory (DFT) to study the repair fragments and their repair processes. The calculations were carried out using Dmol\textthreesuperior{} \citep{17,18} as implemented in Accelrys' Materials Studio (Version 6.0) \citep{19} on the PBE/DNP(3.5)-level \citep{17,20}. The energy convergence criterion for the geometry optimizations was $10^{-5}$ Ha (1 Ha = 1 Hartree = 27.2114 eV = 2625.5 kJ/mol). Each self\-consistent field method (scf) cycle of the geometry optimization steps had a convergence criterion of  $10^{-6}$ Ha. Thermal smearing up to 0.008 Ha was employed for problematic convergence cases. The orbital cutoff was adjusted to 4.6 {\AA}. We included the Grimme DFT-D correction. The reaction energies are calculated for a temperature of 0 K.

\medskip{}

\section{Model system}

\begin{figure}[tb]
\includegraphics{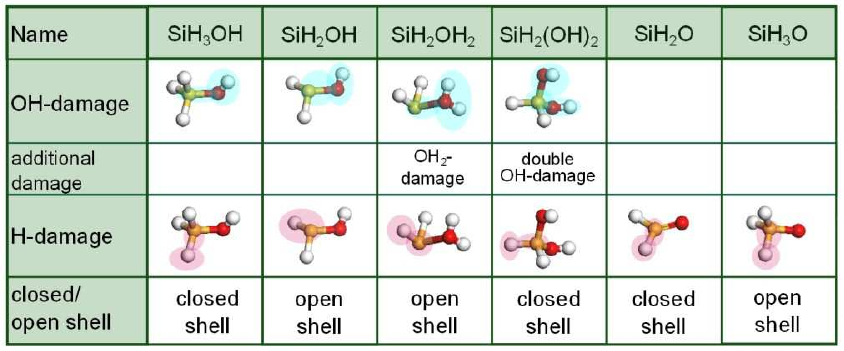}

\protect\caption{(Color online) Overview of the studied ULK-fragments and their damages. The damage
region is highlighted in blue (OH-damage) and pink (H-damage). \label{fig:The-studied-ULK-fragments}}
\end{figure}

To study the repair behavior of the fragments, we composed a small set of ULK-fragments (refer figure \ref{fig:The-studied-ULK-fragments}). The approach to use fragments to study surface reactions has been successfully employed with regard to atomic layer deposition (ALD) processes \citep{21} and was also applied to ULK materials by B{\"o}hm et al. \citep{11,22} .

Opposed to the ULK-fragment used by B{\"o}hm et al, our set of ULK-fragments represents various possible states of silicon and oxygen in the damaged ULK material. Besides closed shell fragments, also open shell fragments were taken into consideration to present under- and overcoordinated silicon and oxygen atoms. With these five ULK-fragments four different damage types can be studied: An H-damage, an OH-damage, a double OH-damage and an $\textrm{O}\textrm{H}_{2}$-damage, as visualized in figure \ref{fig:The-studied-ULK-fragments}. 

All atoms outside the highlighted region in figure \ref{fig:The-studied-ULK-fragments} were fixed during geometry optimizations. The exception is the ULK-fragment $\textrm{Si}\textrm{H}_{2}\textrm{(O}\textrm{H}\textrm{)}_{2}$ whose second hydroxyl group was fixed when representing a single OH-damage. A fast screening of the effectiveness of the considered repair fragments from figure \ref{fig:Overview-repair-reactions} was performed with these five ULK-fragments.

\section{Results and discussion}

\begin{figure}[tb]
\includegraphics{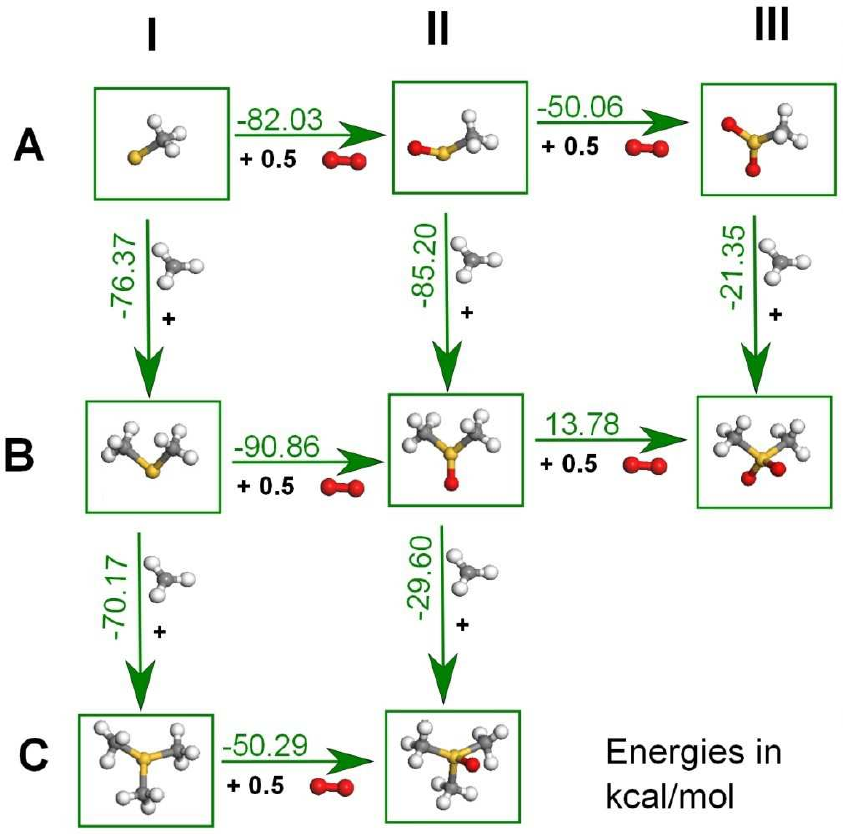}

\protect\caption{(Color online) Selected reactions of the repair fragments. From left to right the
energy for the addition of one oxygen atom to the fragment is shown.
From top to bottom the addition energy for one methyl groups is displayed.
\label{fig:Selected-reactions-of-repair-fragments}}
\end{figure}

There are many silylation precursors available, which could be used
as the basis for the plasma repair process. To study all possible
fragmentation reactions would be far beyond the scope of this paper.
However, today's silylation precursors like \foreignlanguage{american}{octamethylcyclotetrasiloxane}
(OMCTS), \foreignlanguage{american}{bis(dimethylamino)-dimethylsilane}
(DMADMS) and \foreignlanguage{american}{\textcolor{black}{hexamethyldisilazane}}
(HMDS) \citep{23,24} would fragment to at least
one of our considered repair fragments (figure \ref{fig:Overview-repair-reactions}),
as they all consist of at least one silicon atom and one methyl group.
Thus, possible transformation reactions between our considered repair
fragments and the involved reaction energies are shown in a systematic
way in figure \ref{fig:Selected-reactions-of-repair-fragments} . 

The reactions shown are sufficient to demonstrate the predominance
of one repair fragment over the other. In detail, the reaction paths
I to III are highly exothermic and indicate that larger repair fragments
with three methyl groups are preferential. Thus, it is unlikely to
find repair fragments that are prone to damages themselves (see the
group $\textrm{Si}\textrm{O}_{y}\textrm{C}\textrm{H}_{3}$ in Fig.
\ref{fig:Overview-repair-reactions}). Further, reactions A to C demonstrate
that (in most cases) oxygen containing repair fragments are energetically
favored. Finally, we exemplarily study the reaction from A-II to C-I
to investigate the effect of replacing one oxygen atom by two methyl
groups. On the basis of the strongly exothermic reaction energy of
-64.51 kcal/mol, it can be seen that two methyl groups are more effective
than one oxygen atom. In short, under the condition of a methyl rich
atmosphere, $\textrm{Si}\textrm{O}(\textrm{C}\textrm{H}_{3})_{3}$
will be the dominant repair fragment for oxygen containing silylation
precursors, while oxygen-free silylation precursors will lead to $\textrm{Si}(\textrm{C}\textrm{H}_{3})_{3}$.

\begin{figure}[tb]
\includegraphics{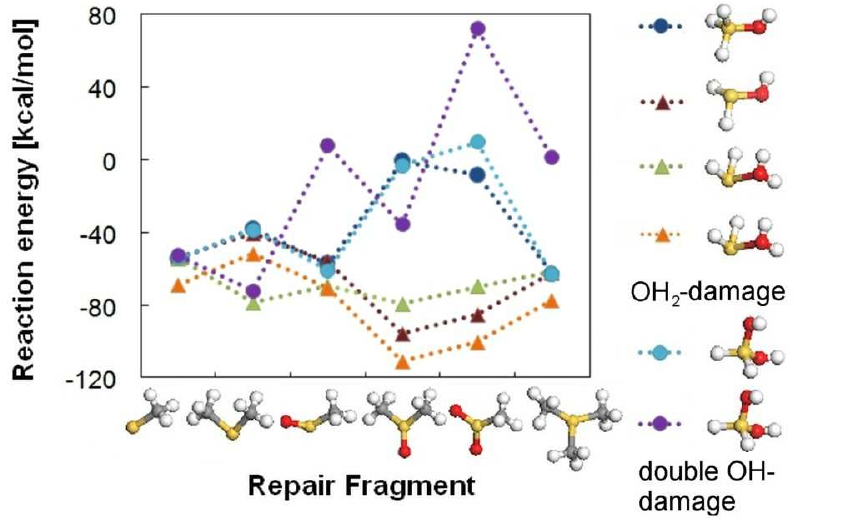}

\protect\caption{(Color online) OH-, double OH- and OH$_{2}$-repair reactions. \foreignlanguage{american}{The
ULK-fragments represented by dots are closed shell fragments, whereas
the open shell ULK-fragments are displayed by triangles in the diagram.
The dotted lines are only a guide for the eyes and show no correlations.}\label{fig:OH-repair-ULK}}
\end{figure}

Next, we analyzed the repair effect of the repair fragments. We first
investigated their reaction with all ULK-fragments from figure \ref{fig:The-studied-ULK-fragments}
that represent an OH-damage. The resulting reaction energies of these
repair processes of an OH-, double OH- and OH$_{2}$-damage are shown
in figure \ref{fig:OH-repair-ULK}.

The main conclusion that can be drawn from the studied OH-repair reactions
is that the reaction energy is strongly dependent on the open shell
or closed shell nature of the ULK-fragments. While the repair fragment
$\textrm{Si}(\textrm{C}\textrm{H}_{3})_{\textrm{3}}$ displays the
best repair effect for closed shell fragments, it is only a mediocre
repair fragment for open shell ULK-fragments. The open shell fragments
instead favor oxygen containing repair fragments, with $\textrm{Si}\textrm{O}(\textrm{C}\textrm{H}_{3})_{\textrm{2}}$
being the best suited fragment for the OH-repair. 

\begin{figure}[t]
\includegraphics{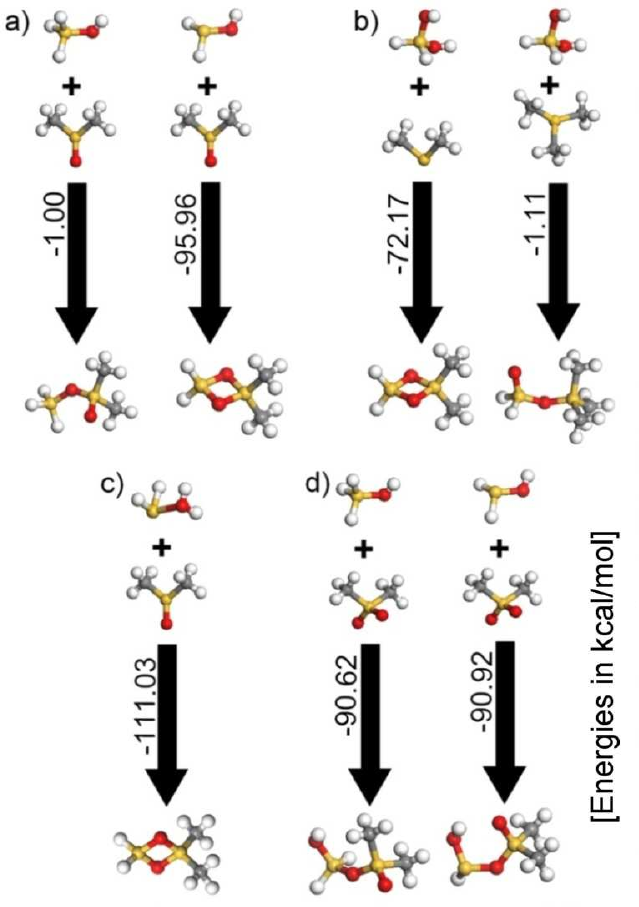}

\protect\caption{(Color online) Exemplary repair reactions for a) OH-damage for \foreignlanguage{american}{$\textrm{Si}\textrm{H}_{3}\textrm{OH}$
(closed shell, left) and $\textrm{Si}\textrm{H}_{2}\textrm{OH}$ (open
shell, right), b) }double OH-damage,\foreignlanguage{american}{ c)}
OH$_{2}$-\foreignlanguage{american}{damage} and d) H-damage for \foreignlanguage{american}{$\textrm{Si}\textrm{H}_{3}\textrm{OH}$
(closed shell, left) and $\textrm{Si}\textrm{H}_{2}\textrm{OH}$ (open
shell, right). The desorbed $\textrm{H}_{2}$ molecule is not displayed.}
\label{fig:Exemplary-repair-reactions}}
\end{figure}

The reason for this different repair tendencies lies in the presence
of one dangling Si-bond in the open shell ULK-fragments. One dangling
Si-bond, together with the oxygen atom of the OH-damage forms an energetically
favorable diamond-shaped SiO-center with the oxygen containing repair
fragments. In the closed shell ULK-fragments, the lack of a dangling
Si-bond together with the saturating hydrogen atoms prevent the formation
of this energetically favored shape (see figure \ref{fig:Exemplary-repair-reactions}a). 

The number of dangling Si-bonds also plays an important role in the
repair of a double OH-damage. Because the repair of one OH-damage
demands exactly one dangling Si-bond in the repair fragment, $\textrm{Si}(\textrm{C}\textrm{H}_{3})_{\textrm{2}}$
is the most effective repair fragment for this damage type. Further,
the two oxygen atoms of the double OH-damage are able to form the
energetically favorable diamond-shaped SiO-center with $\textrm{Si}(\textrm{C}\textrm{H}_{3})_{\textrm{2}}$
as illustrated in figure \ref{fig:Exemplary-repair-reactions}b. For
the majority of the other considered repair fragments, the repair
process is endothermic due to the lack of reactive Si-bonds.

The repair of an OH$_{2}$-damage, on the other hand, is strongly
exothermic for all repair fragments because the formation of water
is energetically favorable. As all other open shell ULK-fragments
of this damage type also prefer oxygen-containg repair fragments over
oxygen-free ones, they lead to the diamond-shaped SiO-center (see
figure \ref{fig:Exemplary-repair-reactions}c).

\begin{figure}[t]
\includegraphics{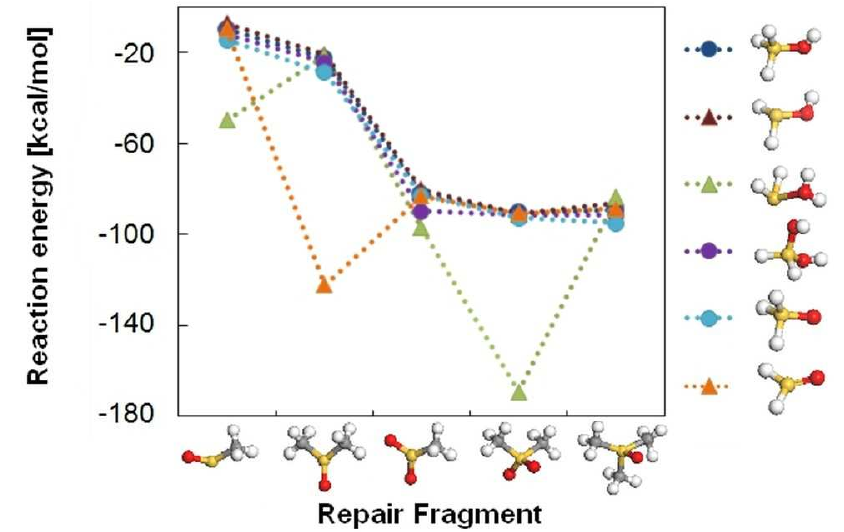}

\protect\caption{(Color online) H-repair of the ULK-fragments. \foreignlanguage{american}{The ULK-fragments
represented by dots are closed shell fragments, whereas the open shell
ULK-fragments are displayed by triangles in the diagram. The dotted
lines are only a guide for the eyes and show no correlations.} \label{fig:H-repair-ULK}}
\end{figure}

Next, we used the ULK-fragments from figure \ref{fig:The-studied-ULK-fragments}
to study the repair of an H-damage and show the results in figure
\ref{fig:H-repair-ULK}. Here, no dependency on the closed shell or
open shell nature is apparent. Figure \ref{fig:Exemplary-repair-reactions}d
illustrates that for these reactions no diamond-shaped SiO-center
is formed because of the fixed position of the saturating OH-group/oxygen
atom in the ULK-fragments. The most effective repair fragments for
the repair of the H-damage are: $\textrm{Si}\textrm{O}_{2}\textrm{C}\textrm{H}_{3},\textrm{ Si}\textrm{O}_{2}(\textrm{C}\textrm{H}_{3})_{\textrm{2}}$
and $\textrm{Si}\textrm{O}(\textrm{C}\textrm{H}_{3})_{\textrm{3}}$.

In a final step we combine our findings of the most effective repair
fragments with the dominant repair fragments that can be gained from
the fragmentation of silylation precursors in a methyl-rich atmosphere.
Based on these studies, we can state that the dominant repair fragments
$\textrm{Si}\textrm{O}(\textrm{C}\textrm{H}_{3})_{\textrm{3}}$ and
$\textrm{Si}(\textrm{C}\textrm{H}_{3})_{3}$ are among the best suited
repair fragments for the OH- and H-damage when a closed shell ULK-fragment
is used which may represent the extended solid best. This means that
the fragments of silylation precursors will be able to successfully
cure H-, OH-, double OH- and $\textrm{O}\textrm{H}_{2}$-damages.
However, the general preference for large repair fragments leads to
a steric hindrance that could prevent the repair of other damages
in close vicinity and thus no full repair of the damaged material
will be achieved.

\textcolor{black}{We find it possible that spatial and electronic
confinements in real pores are likely to prevent the formation of
the energetically favorable diamond-shaped SiO-center. This assumption
is based on the behavior of the closed shell ULK-fragments and the
repair of H-damages. In both cases the space accessible for the oxygen
atom is restricted either by steric hindrance from the H-atoms or
because of its fixed position. This prevents the oxygen atom from
getting in the bonding range of the repair fragment's SiO-group, and
thus no energetically favored diamond-shaped SiO-center can be formed.
This kind of restriction in the movement of oxygen will also occur
in the bulk material, where oxygen is already bound to one silicon
atom \citep{25}.}

\section{Conclusion}

\begin{figure}[h]
\includegraphics{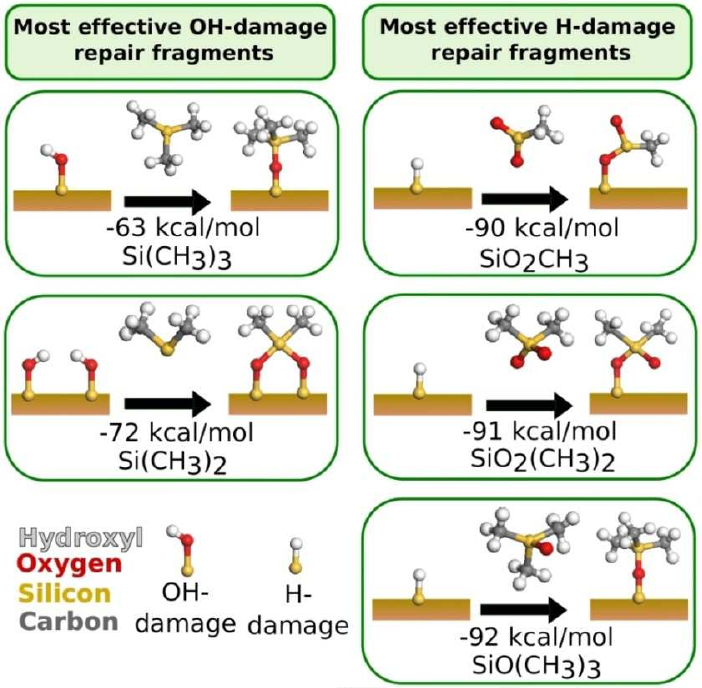}

\protect\caption{(Color online) Summary of the findings. The best suited repair fragments for the
repair of an OH-, double OH, and H-damage are displayed. The energies
are the results from the ULK-fragment $\textrm{Si}\textrm{H}_{2}\textrm{(O}\textrm{H}\textrm{)}_{2}$.
\label{fig:Summary-of-the-findings}}
\end{figure}

\textcolor{black}{We have successfully demonstrated that the plasma
repair process is able to repair both H- and }OH-damages and summarize
our findings in figure \ref{fig:Summary-of-the-findings}. The best
repair fragments for the repair of OH-damages possess the same amount
of dangling bonds as OH-damages that have to be repaired. And oxygen
atoms are needed to repair H-damages.

Thus, we can say that the suggested plasma repair process based on
fragmented silylation precursors is a suitable alternative repair
process for damaged ULK materials. We expect that the final repair
behavior of the plasma repair process will depend on the chosen silylation
precursor. For example, a non-oxygen containing silylation precursor
will be less effective than an oxygen containing one, as it is not
be able to cure H-damages due to the missing oxygen atoms. 

\begin{acknowledgments}
S. Gemming acknowledges funding from the Initiative and Networking
Funds of the President of the Helmholtz Association via the W3 program.
C. Wagner acknowledges the funding by the DFG research unit 1713 \textquotedbl{}Sensoric
Micro- and Nanosystems\textquotedbl{}. 
\end{acknowledgments}

\bibliographystyle{elsarticle-num}

\end{document}